\documentclass[11pt]{article}
\usepackage[utf8]{inputenc}
\usepackage[T1]{fontenc}
\usepackage{amsmath,amsfonts,amssymb}
\usepackage{graphicx}
\usepackage{hyperref}
\usepackage{natbib}
\usepackage{authblk}
\usepackage[letterpaper,top=2cm,bottom=2cm,left=3cm,right=3cm,marginparwidth=1.75cm]{geometry}
\usepackage{booktabs} 
\usepackage{tikz}
\usepackage{appendix}
\usepackage{amsmath}
\usepackage{graphicx}
\usepackage{verbatim}
\usepackage{caption}
\usepackage{multirow}
\usepackage{subfig}
\usepackage[shortlabels]{enumitem}

\graphicspath{{Fig/}}

\newtheorem{theorem}{Theorem}
\newtheorem{definition}{Definition}
\newtheorem{lemma}{Lemma}

\DeclareMathOperator{\doo}{do}

\DeclareMathOperator{\Pa}{Pa}
\DeclareMathOperator{\Ch}{Ch}

\DeclareMathOperator{\An}{An}
\DeclareMathOperator{\De}{De}

\hypersetup{
    colorlinks=true,
    linkcolor=blue,
    filecolor=magenta,      
    urlcolor=cyan,
    citecolor=blue
}

\title{Mediation Analysis in the Presence of Sample Selection Bias with an Application to Disparities in Liver Transplantation Listing}

\author[1]{Zain Khan}
\author[2]{Lynnette Sequeira}
\author[3]{Alexandra T. Strauss}
\author[3,4]{Vedant Jain}
\author[5]{Juliette Dixon}
\author[3]{Eric Moughames}
\author[3]{Tyrus Vong}
\author[6]{Daniel Malinsky}

\affil[1]{Department of Biomedical Engineering, Columbia University}
\affil[2]{Department of Medicine, Johns Hopkins University}
\affil[3]{Division of Gastroenterology and Hepatology, Johns Hopkins University}
\affil[4]{Carle Illinois College of Medicine, University of Illinois}
\affil[5]{Department of Care Management, Johns Hopkins Hospital}
\affil[6]{Department of Biostatistics, Columbia University}

\date{\today}

\begin{document}

\maketitle

\abstract{The study of disparities in the liver transplantation process may focus on quantifying causal effects, particularly the average, direct, or indirect effects of various social determinants of health on being listed as a candidate for transplant. Selection bias arises when the data sample does not represent the target population, defined here as all individuals referred to the transplant clinic. Listing decisions are made for the subset of patients who complete the evaluation process, who may differ systematically from the referred population. There is evidence that selection is associated with patient characteristics that also impact outcomes. Using data only from the selected population may yield biased causal effect estimates. However, incorporating data from the referred population allows for analytic correction. This correction leverages hypothesized causal relationships among selection, the outcome (getting listed), exposures, and mediators. Using directed acyclic graphs (DAGs), we establish graphical conditions under which a reweighted mediation formula identifies effects of interest — direct, indirect, and path-specific effects — in the presence of sample selection. In a clinical case study, we investigate mediated and direct effects of a patient’s socioeconomic position on being listed for transplant, allowing selection to depend on race, gender, age, and other social determinants. }

\textbf{Keywords:} Causal Inference, Liver Transplantation, Mediation Analysis, Selection Bias


\maketitle

\section{Introduction}

A growing body of literature has been examining systematic biases and sources of unfairness in medical decision-making, especially where (partially) automated algorithms are involved in directing patient care \citep{obermeyer2019dissecting,xu2022algorithmic,bhavsar2023defining,chen2023algorithmic}. 
Some approaches to evaluating mechanisms underlying health disparities and unfairness make use of tools from observational causal inference, including the estimation of exposure effects. In particular, many studies leverage mediation analysis: a statistical approach to decomposing the effect of an exposure along ``indirect'' (through an intermediate variable) versus ``direct'' (not through an intermediate variable) causal pathways \citep{vanderweele2016mediation}. More complicated decompositions of causal effects may target path-specific effects (PSEs), which aim to isolate the effects propagating along some specific pathways in settings with multiple intermediate variables between exposure and outcome. 
The goal is often to understand, and potentially mitigate, the different causal mechanisms by which some social determinant of health affects decisions, possibly through some intermediary variables.

Organ transplantation, in particular liver transplantation, is one domain where gender, racial, and socioeconomic position (SEP) disparities have been recognized  \citep{mathur2010racial,nephew2021racial,warren2021racial} and interest in using machine learning for patient prioritization has been growing \citep{khorsandi2021artificial, spann2025role}. This has raised concerns about fairness and possible disparities: algorithms may allocate organ transplants based in part on social determinants of health or measures downstream of these determinants \citep{strauss2023artificial,drezga2023should,dale2024inconsistent}. Associations between a patient's SEP and race and their ``listing" status, or eligibility to receive a liver transplant, have been documented \citep{nephew2021racial,strauss2022multicenter}. Our interest is in distinguishing possible causal pathways from SEP to listing status. In the evaluation phase of the transplant process, social workers perform a ``psychosocial review" and evaluate factors such as the patient's level of social support. This information may in part determine if a patient is listed. For example, social support may be related to SEP: family of patient’s with low SEP may be less able to leave work and be present for their loved one’s medical appointments compared to their high SEP counterparts \citep{strauss2022multicenter}. So, one of the mechanisms by which SEP may affect listing decision is via information used for the psychosocial assessment.

In practice, estimating mediation effects from available clinical data may be affected by \textit{sample selection bias}. The population of interest is comprised of all patients referred to the transplantation clinic (and thus ``at risk'' of being listed for transplant). But the sample available for analysis typically only contains individuals that ``survive'' the entire evaluation process and receive a listing decision. There may be various reasons that some patients do not complete the evaluation process, including the possibility that some patients terminate due to financial hardship, difficulty in attending all requisite appointments (which requires social resources: e.g., time and transportation), or unstable housing. Thus the analysis sample may not be entirely representative of the referred population. However, it may be possible to analytically adjust for selection bias in this setting, provided that certain baseline data on the target population are available.

Causal graphical models, especially directed acyclic graphs (DAGs) and their generalizations, have been useful for representing assumptions about complex data-generating processes and evaluating potential sources of bias that may impact the estimation of causal effects from observational data. Identification theory based graphical models has been applied to questions in both mediation analysis and to address selection bias, but not simultaneously.

Our contributions in this manuscript are the following. First we provide sufficient graphical conditions to identify mediation effects and PSEs under selection bias. We provide corresponding formulas for mediation effects and PSEs in the presence of selection bias for when these identification conditions hold. We apply this theory to clinical data from a large urban medical system, estimating direct and indirect effects of SEP on transplant listing decision relative to an important mediator, the outcome of psychosocial review (which is a precursor to listing decision). We contrast our selection adjusted estimated mediation effects with estimates that make no adjustment for selection bias.

\section{Background}


\subsection{Graphical Models and Mediation}

We use graphical models along with the potential outcomes notation to describe causal effects and selection/confounding biases. For relevant background see \citep{pearl2009causality,hernan2016using}. A graph $G=(V,E)$ consists of a set of vertices $V$ and edges $E$, and we allow that $E$ may include both directed ($\rightarrow$) and bidirected ($\leftrightarrow$) edges. We assume $G$ is acyclic here, and so $G$ is formally an acyclic directed mixed graph (ADMG). A directed edge $V_i \rightarrow V_j$ represents that $V_i$ is ``direct'' cause of $V_j$ and a bidirected edge $V_i \leftrightarrow V_j$ represents that $V_i$ and $V_j$ are associated due to a shared unmeasured common cause. In the special case with only directed (no bidirected) edges, $G$ is a directed acyclic graph (DAG). For a vertex $V_i \in V$, we use common graph theoretic definitions such as $\Ch(V_i), \Pa(V_i), \De(V_i), \An(V_i)$ to denote the children, parents, descendants, and ancestors, respectively, of $V_i$ in $G$. A vertex $V_k$ lying on a path from $V_i$ to $V_j$ is called a collider on that path if both edges incident to $V_k$ have arrowheads at $V_k$, e.g., $\rightarrow V_k \leftarrow$, $\leftrightarrow V_k \leftrightarrow$, $\rightarrow V_k \leftrightarrow$, or $\leftrightarrow V_k \leftarrow$. A path from $V_i$ to $V_j$ is blocked by $Z \subseteq V\setminus \{V_i, V_j\}$ if there exists a non-collider on the path that is in $Z$ or if there exists a collider on the path that is neither in $Z$ and nor an ancestor of $Z$. We say $X$ is m-separated from $Y$ given $Z$ in $G$ if every path from an element of $X$ to an element of $Y$ is blocked by $Z$ in $G$. d-separation is the special case of m-separation when $G$ is a DAG. We use the term \emph{causal path} from $V_i$ to $V_j$ to refer to a directed path from $V_i$ to $V_j$. We use $G_{\overline{X}}$ and $G_{\underline{X}}$ to denote $G$ with directed edges into $X$ removed and directed edges out of $X$ removed, respectively.

We will primarily be concerned with disjoint subsets of variables $X, M, Y \subseteq V$, where $X$ is a set of exposures, $M$ is a set of post-exposure intermediary variables, and $Y$ is a set of outcome variables. We refer to the potential outcome (a.k.a.\ counterfactual) random variable $Y(x)$ as the value $Y$ would take if the random variable $X$ were set to value $x$. The distribution $p(Y(x))$ can also be written as $p(y\mid \doo(x))$ using Pearl's do-notation \citep{pearl2009causality}. The average (or total) causal effect of $X$ on $Y$ is written $E[Y(x)-Y(x')]$ on the mean-difference scale, for two different values $x, x'$. We may consider potential outcomes with multiple arguments: e.g., the counterfactual $Y(x, m)$ refers to the value $Y$ would take if $X$ is set to $x$ and $M$ is set to $m$. Quantities in mediation analysis may refer to nested potential outcomes, such as $Y(x, M(x'))$, which represents the value $Y$ would take if $X$ is set to $x$ but $M$ behaves as if $X$ were set to another value $x'$.

The decomposition 
\begin{align}
\begin{split}
E[Y(x) - Y(x')] = \hspace{1mm} &E[Y(x) - Y(x, M(x'))] \hspace{1mm} + \\
  &E[Y(x, M(x')) - Y(x')]
\end{split}
\end{align}
splits the total effect of exposure into the natural indirect effect (NIE) $E[Y(x) - Y(x, M(x'))]$ and natural direct effect (NDE) $E[Y(x, M(x')) - Y(x')]$ \citep{robins1992identifiability,pearl2022direct,vanderweele2013three,vanderweele2016mediation}. 
The NIE quantifies the effect of exposure $X$ on the outcome $Y$ through the intermediate variable $M$ and the NDE quantifies the effect along all other pathways (not through $M$).

The total effect of exposure may also be represented on a risk ratio scale, useful when the outcome $Y$ is discrete, which is written $E[Y(x)] / E[Y(x')]$ for two different values $x, x'$. The decomposition on the risk ratio scale 
\begin{align}
\begin{split}
\frac{E[Y(x)]}{E[Y(x')]} = \hspace{1mm} & \frac{E[Y(x)]}{E[Y(x, M(x'))]} \hspace{1mm} \times \frac{E[Y(x, M(x'))]}{E[Y(x')]}
\end{split}
\end{align}
similarly splits the total effect of exposure into the natural indirect effect (NIE-RR) $\frac{E[Y(x)]}{E[Y(x, M(x'))]}$ and natural direct effect (NDE-RR) $\frac{E[Y(x, M(x'))]}{E[Y(x')]}$ \citep{vanderweele2014mediation}.

When there is more than one mediator or set of pathways of interest, path-specific effects generalize mediation effects by considering the effects of exposure along an arbitrary set of proper causal paths. 
A proper casual path between from $X$ to $Y$ in a graph $G$ is a causal path that does not intersect $X$ except at the source of the path.
Let $\pi$ be a set of proper causal paths in graph $G$. The potential outcome $V_i(\pi, x, x')$ is defined by setting $X$ to $x$ for the purposes of paths in $\pi$ and $X$ to $x'$ for all proper causal paths not in $\pi$. 
$V_i(\pi, x, x')$ is said to be edge inconsistent if potential outcomes of the form $V_j(x_k,\dots)$ and $V_j(x_k',\dots)$ occur in $V_i(\pi, x, x')$, otherwise it is said to be edge consistent. Edge inconsistent quantities are not generally identified \citep{avin2005identifiability} so we will only consider edge consistent quantities in this work. 

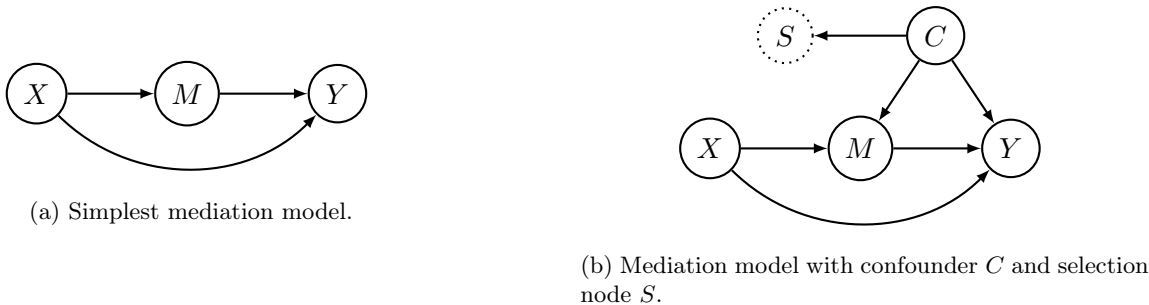
\begin{figure}[h!]
    \centering
    \subfloat[Simplest mediation model.\label{fig:simple_mediation}]{%
        \begin{minipage}{0.45\textwidth}
            \centering
            \begin{tikzpicture}[->, >=latex, thick]
                \node (X) [circle, draw] at (0, 0) {$X$};
                \node (M) [circle, draw] at (2, 0) {$M$};
                \node (Y) [circle, draw] at (4, 0) {$Y$};

                \draw[->] (X) -- (M);
                \draw[->] (M) -- (Y);
                \draw[->] (X) to[bend right=45] (Y);
            \end{tikzpicture}
        \end{minipage}
    }%
    \hfill 
    \subfloat[Mediation model with confounder \( C \) and selection node \( S \).\label{fig:confounded_mediation}]{%
        \begin{minipage}{0.45\textwidth}
            \centering
            \begin{tikzpicture}[->, >=latex, thick]
                \node (S) [circle, draw, dotted] at (1, 1.5) {$S$};
                \node (C) [circle, draw] at (3, 1.5) {$C$};
                \node (X) [circle, draw] at (0, 0) {$X$};
                \node (M) [circle, draw] at (2, 0) {$M$};
                \node (Y) [circle, draw] at (4, 0) {$Y$};

                \draw[<-] (S) -- (C);
                \draw[->] (C) -- (M);
                \draw[->] (C) -- (Y);
                \draw[->] (X) -- (M);
                \draw[->] (M) -- (Y);
                \draw[->] (X) to[bend right=45] (Y);
            \end{tikzpicture}
        \end{minipage}
    }
    \caption{Examples of mediation models}
    \label{fig:mediation_models}
\end{figure}

We depict a simple mediation model in Figure \ref{fig:mediation_models} and a more complex mediation model with multiple mediators in Figure \ref{fig:complex_mediation_models}. In the latter model, one PSE of interest may be along the paths highlighted in red, with corresponding counterfactual $Y(\pi,x,x') = Y(x',M_1(x),M_2(x', M_1(x)))$ for $\pi = \{X \rightarrow M_1 \rightarrow Y, X \rightarrow M_1 \rightarrow M_2 \rightarrow Y\}$. Note that with a single mediator $M$, the direct or indirect effects studied in mediation analysis are special cases of path-specific effects. For example, in Figure \ref{fig:mediation_models} (a) and (b), by choosing $\pi = \{ X \rightarrow M \rightarrow Y \}$, the $\pi$-specific effect is the indirect effect.

\subsection{Confounding and Selection Bias}

One possible source of bias in observational studies is bias from confounding. Given an assumed graphical representation of the causal relationships among study variables, confounding bias may be addressed analytically if some subset of the measured covariates comprise a valid adjustment set.  
\begin{definition} 
\textbf{Adjustment Set}. Given a causal graph $G$, and variables $X, Y \subseteq V$, the set of variables $Z \subseteq V\setminus\{X,Y\}$ is called an adjustment set for the effect of $X$ on $Y$ if for all possible joint distributions, $p(v)$, the following holds:
\begin{equation}
p(Y(x)) = \sum_z p(y \mid x, z) \; p(z)
\label{adj}
\end{equation}
\end{definition}

The identification formula (\ref{adj}) is referred to as the ``adjustment formula'' or ``g-formula.'' Given a graph $G$, there is well-known graphical criterion for determining whether some set of covariates is a valid adjustment set: the backdoor criterion. To satisfy the backdoor criterion $Z$ must satisfy two conditions: (a) $\forall Z_i \in Z$, $Z_i \not \in \De(X)$ and (b) $Z$ blocks all non-causal paths between $X$ and $Y$ \citep{pearl2009causality}. Condition (b) can be equivalently stated as the requirement that $Z$ m-separates $X$ and $Y$ in the proper backdoor graph between $X$ and $Y$, $G^{pbd}_{X, Y}$ \citep{van2014constructing}.

\begin{definition}
\textbf{Proper Backdoor Graph}. Given a causal graph $G$, and disjoint subsets $X, Y$ of $V$, the proper backdoor graph, $G^{pbd}_{X, Y}$, removes the first edge along every proper causal path from $X$ to $Y$.
\end{definition}

The backdoor criterion thus provides a graphical tool to determine which sets of covariates, if any, are sufficient to control for confounding bias.

Another important source of possible bias is (sample) selection bias. This type of bias arises from non-random selection into the analysis sample such that certain units, e.g.\ candidates for transplantation, are differentially less likely to be included in the analysis and the likelihood of inclusion is related to important covariates, such as demographics or socioeconomic position \citep{strauss2022multicenter,strauss2023critical}. This selection may happen unintentionally. 

To graphically represent the selection phenomenon, one may augment a causal graph with a binary selection node, denoted as $S$. Observed (selected) units will have the value $S=1$ while unobserved units will have $S=0$. This selection node is either a child of or bidirected-connected to some other study variables. For example if older patients were less likely to be included in the analysis sample for liver transplant evaluation (having not completed the evaluation process), $S$ would be a child of the node that corresponds to age.

To analytically correct for the effect of selection bias in practice, one may use an approach closely related to confounding adjustment. Graphical identification results for causal inference in the presence of selection have been developed by several authors \citep{hernan2004structural,bareinboim2012controlling,bareinboim2015recovering,correa2017causal,correa2018generalized,bareinboim2022recovering,mathur2025simple}.
One approach combines covariate adjustment with ``external'' data (i.e., data from the full population of interest, not affected by the selection mechanism) on some covariates. For example, in some studies certain baseline covariates such as demographic variables are recorded for the full target population, even if most study variables (including outcomes) are available only for the selected. Using this information, \citet{correa2018generalized} have introduced an adjustment formula that identifies causal effects from selected samples. Corresponding to this adjustment formula, there is a graphical criterion that \citet{correa2018generalized} call the Generalized Adjustment Criterion (GAC).\footnote{The authors present several adjustment criteria; the one here is their GAC Type 3.} In the following, $Z$ is a set of covariates and $Z^T \subseteq Z$ is the subset of covariates for which ``external'' data (data from the full target population) is available.

\begin{definition}
\textbf{Generalized Adjustment Criterion}. Given a causal graph $G$ augmented with selection node $S$, disjoint sets of variables $Z, X, Y \subseteq V$, and a set $Z^T \subseteq Z$; $(Z, Z^T)$ is an admissible pair relative to $X, Y$ in $G$ if
\begin{itemize}
    \item[1. ] No element in $Z$ is a descendant in $G_{\overline{X}}$ of any $W \not \in X$ lying on a proper causal path from $X$ to $Y$
    \item[2. ] All backdoor paths between $X$ and $Y$ are blocked by $Z$ and $S$, i.e., $(X \perp Y \mid Z, S)_{G^{pbd}_{X, Y}}$
    \item[3. ] $Z^T$ m-separates $Y$ from $S$ in the proper backdoor graph, i.e., $(Y \perp S \mid Z^T)_{G^{pbd}_{X, Y}}$
 \end{itemize}
\end{definition}

\citet{correa2018generalized} proved that a pair $(Z, Z^T)$ is admissible according to the criterion above if and only if the following formula holds:
\begin{equation}
p(Y(x)) = \sum_z p(y \mid x, z, S=1) \; p(z \setminus z^T \mid z^T, S=1) \; p(z^T)
\label{eqn:GAC3}
\end{equation}

This can be seen as a version of the backdoor adjustment formula (\ref{adj}) extended to the setting with selection bias and some external data. On the right-hand side, all quantities except one appear conditional on $S=1$, indicating that these refer to distributions in the selected sample. The last quantity, the distribution of $Z^T$, does not condition on $S=1$ because information on $Z^T$ is assumed to be available from the full target population.

\section{Identification of Path Specific Effects under Selection Bias}


The identification of mediation quantities (including direct/indirect effects and path-specific effects) depends on additional assumptions beyond those encoded directly in the graphical model $G$. In particular, there are three distinct causal models that have figured prominently in discussions of mediation analysis: the NPSEM-IE (nonparametric structural equations model with independent errors a.k.a.\ the ``multiple worlds model''), the FFRCISTG model of Robins \citep{robins1986new}, and the ``extended graph'' or ``split-treatment'' approach first proposed by \citet{robins2010alternative} and then generalized and expanded by \citet{malinsky2019potential,didelez2019defining,robins2022interventionist}. See \citet{robins2022interventionist} for extensive discussion of the relationship among these approaches. We adopt a version of the ``extended graph'' approach here. (Our results also apply under the NPSEM-IE model.)

\subsection{Mediation and Path-Specific Effects with Extended Graphs}

Given a causal graph $G$, outcome $Y$, exposure $X$, and set of mediators $M$, mediation and path specific effects can be formulated using a construction called the extended graph of $G$, denoted as $G^e$ as presented in \citet{malinsky2019potential}. $G^e$ is identical to $G$ except that on every proper causal path from $X$ to $Y$, the path will be extended with a new child of $X$. That is, the first edge out of $X$ on any path of the form $X \rightarrow Y$ or $X \rightarrow M \rightarrow ... \rightarrow Y$  is replaced with $X \rightarrow X_y^e \rightarrow Y$ and $X \rightarrow X_{m}^e \rightarrow M \rightarrow ... \rightarrow Y$, respectively. The set of extended nodes in $G^e$ introduced via this construction is denoted as $X^e$. The edges from $X$ to $X^e$ are understood to represent deterministic relationships: the extended children take on the same values as their parent nodes. Refer to Figure \ref{fig:extended_mediation_models} for the extended variants of the causal graphs presented in Figure \ref{fig:mediation_models}. (Note: in earlier work the extended graph included extended nodes for all edges out of $X$, but our restriction here to proper causal paths is simpler and sufficient for the quantities in this work.)

\begin{figure}[h!]
    \centering
    \subfloat[Simplest mediation model with extended nodes \(X^e_m\) and \(X^e_y\).\label{fig:extended_simple}]{%
        \begin{minipage}{0.45\textwidth}
            \centering
            \begin{tikzpicture}[->, >=latex, thick, node distance=2cm]
                \tikzset{node style/.style={circle, draw, minimum size=1cm, font=\small, align=center}}
                \node (X) [node style] at (0, 0) {$X$};
                \node (XeM) [node style] at (2, 0) {$X^e_m$};
                \node (M) [node style] at (4, 0) {$M$};
                \node (XeY) [node style] at (2, -1.5) {$X^e_y$};
                \node (Y) [node style] at (6, 0) {$Y$};
                \draw[->] (X) -- (XeM);
                \draw[->] (XeM) -- (M);
                \draw[->] (M) -- (Y);
                \draw[->] (X) -- (XeY);
                \draw[->] (XeY) -- (Y);
            \end{tikzpicture}
        \end{minipage}
    }%
    \hfill 
    \subfloat[Mediation model with confounder \( C \), selection node \( S \), and extended nodes \(X^e_m\) and \(X^e_y\).\label{fig:extended_confounded}]{%
        \begin{minipage}{0.45\textwidth}
            \centering
            \begin{tikzpicture}[->, >=latex, thick, node distance=2cm]
                \tikzset{node style/.style={circle, draw, minimum size=1cm, font=\small, align=center}}
                \node (S) [node style, dotted] at (3, 2) {$S$};
                \node (C) [node style] at (5, 2) {$C$};
                \node (X) [node style] at (0, 0) {$X$};
                \node (XeM) [node style] at (2, 0) {$X^e_m$};
                \node (M) [node style] at (4, 0) {$M$};
                \node (XeY) [node style] at (2, -1.5) {$X^e_y$};
                \node (Y) [node style] at (6, 0) {$Y$};
                \draw[<-] (S) -- (C);
                \draw[->] (C) -- (M);
                \draw[->] (C) -- (Y);
                \draw[->] (X) -- (XeM);
                \draw[->] (XeM) -- (M);
                \draw[->] (M) -- (Y);
                \draw[->] (X) -- (XeY);
                \draw[->] (XeY) -- (Y);
            \end{tikzpicture}
        \end{minipage}
    }
    \caption{Extended causal graphs for the mediation models in Fig. \ref{fig:mediation_models}.}
    \label{fig:extended_mediation_models}
\end{figure}
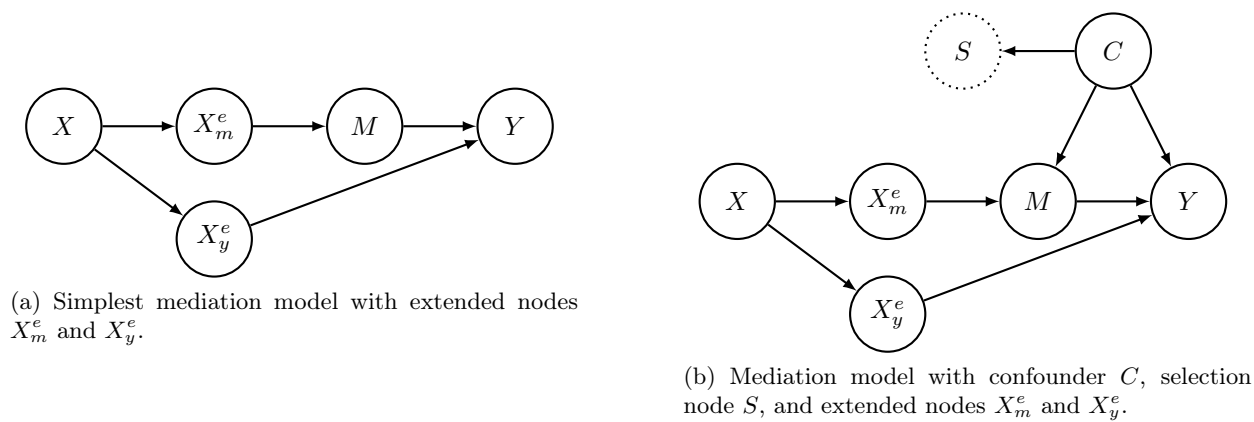

\begin{figure}[h!]
    \centering
    \subfloat[A mediation model with two mediators, $M_1$ and $M_2$. Red edges indicate a set of paths of possible interest, $\pi=\{X \rightarrow M_1 \rightarrow Y, X \rightarrow M_1 \rightarrow M_2 \rightarrow Y\}$.\label{fig:two_mediators_simple}]{%
        \begin{minipage}{0.45\textwidth}
            \centering
            \begin{tikzpicture}[->, >=latex, thick, node distance=1.5cm and 2cm]
                \tikzset{node style/.style={circle, draw, minimum size=1cm, font=\small, align=center}}
                \node (X) [node style] at (0, 0) {$X$};
                \node (M1) [node style] at (3, 0.75) {$M_1$};
                \node (M2) [node style] at (3, -0.75) {$M_2$};
                \node (Y) [node style] at (6, 0) {$Y$};
                \draw[->, red] (X) -- (M1); 
                \draw[->, red] (M1) -- (M2);
                \draw[->] (X) -- (M2);
                \draw[->, red] (M2) -- (Y);
                \draw[->] (X) to[bend right=45] (Y);
                \draw[->, red] (M1) to (Y);
            \end{tikzpicture}
        \end{minipage}
    }%
    \hfill 
    \subfloat[Extended node expansion of (a).\label{fig:two_mediators_extended}]{%
        \begin{minipage}{0.45\textwidth}
            \centering
            \begin{tikzpicture}[->, >=latex, thick, node distance=1.5cm and 2cm]
                \tikzset{node style/.style={circle, draw, minimum size=1cm, font=\small, align=center}}
                \node (X) [node style] at (0, 0) {$X$};
                \node (XeM1) [node style] at (2, 0.75) {$X^e_{m_1}$};
                \node (M1) [node style] at (4, 0.75) {$M_1$};
                \node (XeM2) [node style] at (2, -0.75) {$X^e_{m_2}$};
                \node (XeY) [node style] at (3, -1.75) {$X^e_{Y}$};
                \node (M2) [node style] at (4, -0.75) {$M_2$};
                \node (Y) [node style] at (6, 0) {$Y$};
                \draw[->, red] (X) -- (XeM1);
                \draw[->, red] (XeM1) -- (M1); 
                \draw[->] (X) -- (XeM2);
                \draw[->] (X) to[bend right=30] (XeY);
                \draw[->, red] (M1) -- (M2);
                \draw[->] (XeM2) -- (M2);
                \draw[->, red] (M1) -- (Y);
                \draw[->, red] (M2) -- (Y);
                \draw[->] (XeY) to[bend right=30] (Y);
            \end{tikzpicture}
        \end{minipage}
    }%
    
    \caption{Example mediation models with multiple mediators.}
    \label{fig:complex_mediation_models}
\end{figure}
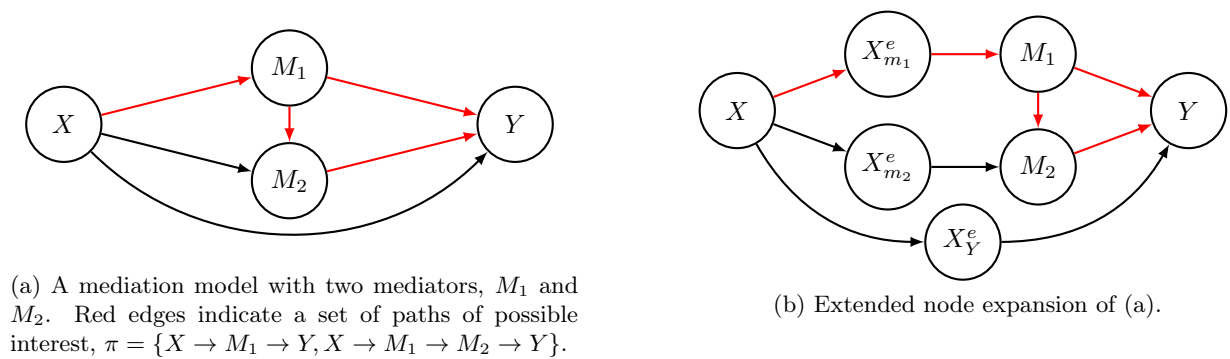

Given the close relationship between $G^e$ and $G$, we have the following result: an adjustment set $(Z, Z^T)$ that is valid to address confounding and selection given $G$ is also valid when considering $X^e$ as the exposure in the extended graph $G^e$. This will be useful for establishing an analagous statement for the mediation setting.

\begin{theorem}
\label{theorem:GACiff}
$(Z, Z^T)$ is an admissible pair relative to $X, Y$ in $G$ if and only if $(Z, Z^T)$ is an admissible pair relative to $X^e, Y$ in $G^e$. 
\end{theorem}

The proof makes use of the Generalized Adjustment Criterion above and is deferred, with all other proofs, to the supplementary material.


Extended graphs enable a redefinition of mediation and path-specific quantities in terms of joint interventions (do-interventions) on extended nodes. That is, we can replace complex nested counterfactual quantities with quantities that resemble plain interventional distributions. For example, in Figure \ref{fig:complex_mediation_models} (a), the path-specific effect of interest (the effect along all paths through $M_1$) involves the nested counterfactual $Y(x', M_1(x), M_2(x', M_1(x)))$, which we can equivalently express as a joint intervention that sets $X^e_{m_1}$ to ``active'' value $x$ and all remaining extended nodes to the reference value $x'$: 

\begin{equation*}
\begin{aligned}
E[Y(x', M_1(x), M_2(x', M_1(x)))]
=  E[Y \mid \doo(X^e_y = x', X^e_{m_1} = x, X^e_{m_2} = x')]
\end{aligned}
\label{eqn:med-example}
\end{equation*}

\subsection{Selected Mediation Formula}



\citet{pearl2022direct} introduced a popular \textit{mediation formula} for the purposes of calculating direct and indirect effects. This formula expresses the mean nested counterfactual $E[Y(x, M(x'))]$ -- the key ingredient of the NIE and NDE -- as a function of the observed data distribution (in the setting with no sample selection bias).

\begin{equation}
\begin{aligned}
E[Y(x,M(x'))]
= &\sum_{z,m} E[Y \mid X=x, m, z] \; p(m \mid X=x', z) \; p( z )    \\
\end{aligned}
\label{eqn:med-pearl}
\end{equation}

First we consider the simplest setting with only a single mediatior between $X$ and $Y$. A commonly-stated sufficient condition for the validity of (\ref{eqn:med-pearl}), in addition to the assumption that $Z$ blocks all backdoor paths between $Y$ and $X$, is the ``cross-world'' independence assumption $Y(x, m) \perp M(x') \mid Z$. This assumption is called ``cross-world'' because it simultaneously evokes the ``world'' where $Y$ behaves as if $X=x$ and the ``world'' where $M$ behaves as if $X=x'$ \citep{robins2010alternative, richardson2013single}. Alternatively, with the introduction of extended nodes and the corresponding extended graphical formalism, the relevant assumption may be stated with respect to the extended nodes $X^e_y, X^e_m$ -- that is, one considers a joint intervention that sets $X^e_y=x^e_y$ and sets $X^e_m=x'^e_m$. If $Y(x^e_y, m) \perp M(x'^e_m) \mid Z$ (and $Z$ blocks all backdoor paths between $Y$ and $X^e_y, X^e_m$), then (\ref{eqn:med-pearl}) holds. We combine this extended graph formalism with the Generalized Adjustment Criterion to present our main theoretical result, which extends the mediation formula to settings with sample selection.

In the setting with selection bias, we propose the \textit{selected mediation formula}:

\begin{equation}
\begin{aligned}
E[Y(x, M(x'))] 
= &\sum_{z,m} E[Y \mid X=x, m, z, S=1] \; p(m \mid X=x', z, S=1) \;  p( z \setminus z^T \mid z^T, S=1) \; p(z^T )
\end{aligned}
\label{eqn:med}
\end{equation}

\begin{theorem}

Given a causal graph $G$ augmented with selection node $S$ and a single mediator $M = \Ch(X) \cap \Pa(Y)$, the selected mediation formula (Eq.\ \ref{eqn:med}) holds if: 

\begin{itemize}
    \item[1. ] The GAC conditions hold given $(Z, Z^T )$ in $G$ for the total effect of $X$ on $Y$, and
    \item[2. ] All backdoor paths between $M$ and $Y$ are blocked by $Z$ and $S$, i.e., $(M \perp Y \mid Z, S)_{G^{pbd}_{(X, M), Y}}$
 \end{itemize}

\label{theorem:mediationadmg}
\end{theorem}
\noindent In the more general setting with multiple mediators, we can state a similar result for arbitrary path-specific effects. 

\begin{theorem}
Given a causal graph $G$ augmented with selection node $S$, the edge-consistent $\pi$-specific effect of $X$ on $Y$, written $p(Y(\pi,x,x'))$, is identified by the adjustment formula below if:

\begin{itemize}
    \item[1. ] The GAC conditions hold given $(Z, Z^T )$ in $G$ for the total effect of $X$ on $Y$, and
    \item[2. ] All backdoor paths between $M_i \in M$ and $Y$ are blocked by $Z$ and $S$, i.e., $(M_i \perp Y \mid Z, S)_{G^{pbd}_{(X, M_i), Y}}$ for all $M_i$, where $M$ contains all nodes along proper causal paths from $X$ to $Y$.
\end{itemize}

\begin{equation}
\begin{aligned}
p(Y(\pi,x,x'))
=\\ &\sum_{z, m} p(y \mid x \cap \Pa_y^{\pi}, x' \cap \Pa_y^{\bar{\pi}}, m, z, S=1) \\
  &\times\; \prod_{i=1}^{ |M| } p(m_i \mid  x \cap \Pa_{m_i}^{\pi}, x' \cap  \Pa_{m_i}^{\bar{\pi}}, \Pa^{M}(M_i), z, S=1) \\
  &\times\; p(z \setminus z^T \mid z^T, S=1) p(z^T) \\
\end{aligned}
\end{equation}
\noindent Here $\Pa^{M}(M_i) \equiv \Pa(M_i) \cap M$. The notation $x \cap \Pa_y^{\pi}$ denotes that parents of $Y$ in $X$ on $\pi$ are set to values in $x$ while $x' \cap \Pa_y^{\bar{\pi}}$ denotes that parents of $Y$ in X not on $\pi$ are set to values in $x'$.

\label{theorem:pseadjustment}
\end{theorem}

This can be seen as a fusion of what \cite{shpitser2016causal} call the ``edge g-formula'' with the GAC approach to adjusting for selection bias with external data. In both results above, the second condition rules out certain pathways between mediators and outcomes. This is closely related to the prohibition of ``recanting witnesses'' and ``recanting districts'' in the identification theory of path-specific effects \citep{avin2005identifiability,shpitser2013counterfactual}. A recanting district is a subset of variables in $G$ that precludes identification of path-specific effects by inducing associations that makes it impossible to decompose the joint distribution into components that isolate effects along different pathways. (A recanting witness is a special case of a recanting district when there are no unmeasured confounders, i.e., the model is a DAG rather than an ADMG.) The formal definitions are deferred to the supplementary materials, where we prove an auxiliary lemma to show that assuming the second condition in above is sufficient to rule out recanting districts for the effects of interest.

Given these identification results, in applied settings that are judged to meet the requisite conditions we may adapt popular estimation procedures for direct, indirect, and path-specific effects with inverse probability of selection weights. Following the approach proposed in \citet{correa2018generalized}, define weights based on the external data:

\begin{equation}
\begin{aligned}
w_i = \frac{P(S=1)}{P(S=1 \mid z_i^T)}.
\end{aligned}
\label{eqn:weights}
\end{equation}
These weights may be estimated using a parametric model for the conditional probability of selection given covariates in $Z^T$ and then combined with existing estimators for mediation effects. To illustrate, we use logistic regression to estimate the weights and incorporate these into counterfactual imputation estimators for the NDE and NIE \citep{vansteelandt2012imputation}.

\section{Simulation Study}

To illustrate the proposed selection mediation formula and how it correctly mitigates sample selection bias, we present a a brief simulation study. The data generating process is designed based on Figure \ref{fig:mediation_models} (b) where a mediator-outcome confounder determines selection into the sample. The DGP is as follows: 

\begin{itemize}
    \item[] $X \sim \text{Bernoulli}(0.5)$
    \item[] $C \sim \mathcal{N}(0, 1)$
    \item[] $M = 1.0 \cdot X + 1.0 \cdot C + \epsilon_M$, where $\epsilon_M \sim \mathcal{N}(0, 1)$
    \item[] $Y = 0.5 \cdot X + 1.0 \cdot M + 2.0 \cdot (M \cdot X) + 0.5 \cdot C + \epsilon_Y$, where $\epsilon_Y \sim \mathcal{N}(0, 1)$
\end{itemize}
Selection bias was introduced post-data generation based on the confounder $C$. The probability of being selected into the sample, $S = 1$, was modeled using a logistic regression with $C$: $\operatorname{logit } P(S =1 \mid C) = \beta_S \cdot C$. The parameter $\beta_S$ was varied across values in the range $[0, 2]$ to represent different levels of confounding-induced selection bias. A value of $\beta_S = 0$ indicates random sample selection and higher values of $\beta_S$ indicate stronger selection bias. 

\begin{figure}[h!]
    \centering
    \includegraphics[width=0.5\textwidth]{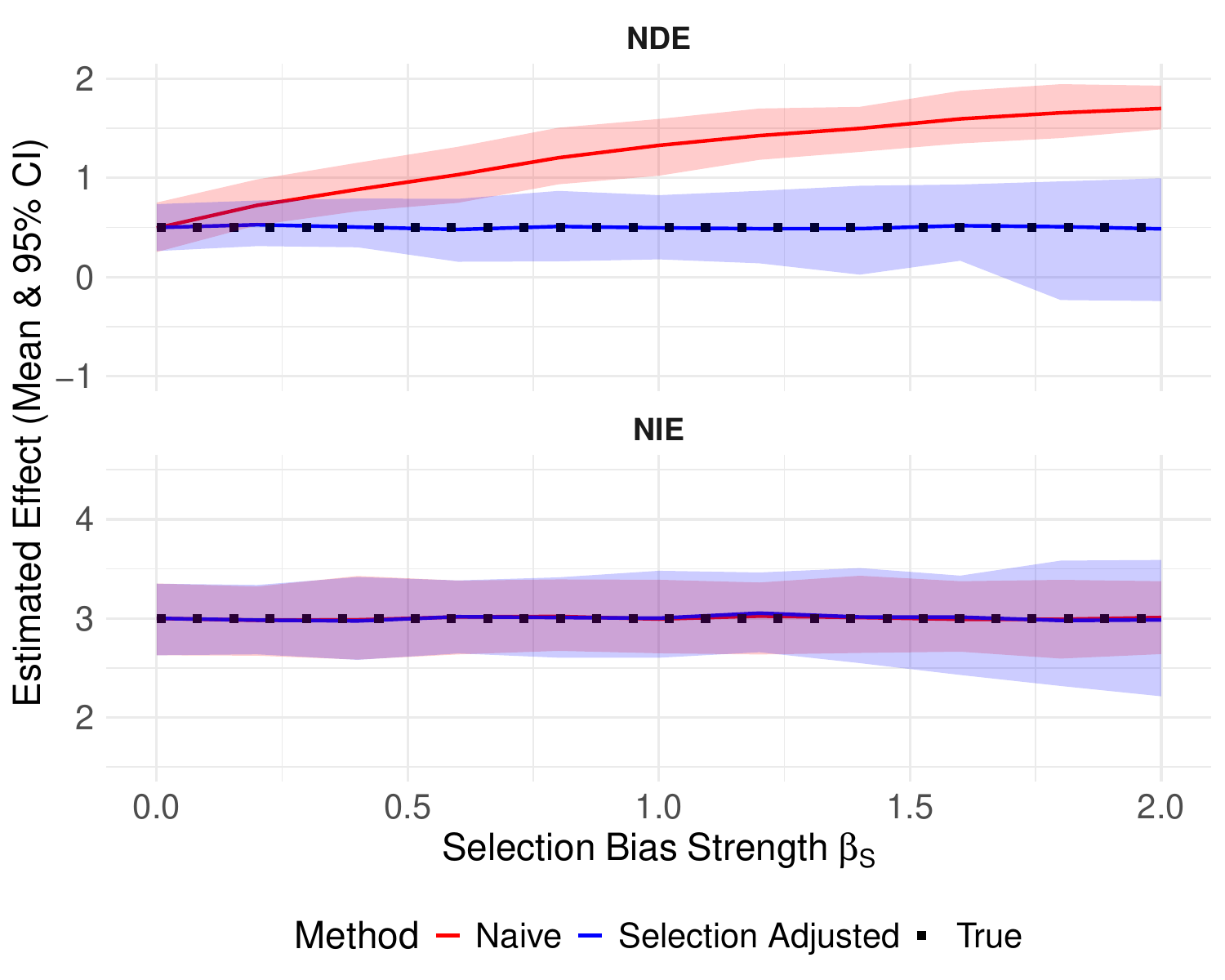} 
    \caption{Comparison of mediation effect estimates across selection bias strength levels. The red line represents the naive (standard) method, the blue line represents the selection adjusted method, and the dashed black line indicates the true effect which is constant. The plots show the mean effect estimates and 95\% confidence intervals.}
    \label{fig:mediation_comparison}
\end{figure}

For every value of $\beta_S$, we estimated the NIE and NDE using both a standard (``naive'') estimator and the selection adjusted estimator, which incorporates inverse probability weights based on estimated selection probabilities. This was repeated $500$ times for datasets of size $n_{0} = 10,000$. After inducing dropout due to selection in this sample, we randomly sampled a subset of $n = 1000$ unique rows with which we estimated the NIE and NDE. This was done to enable fair comparisons at a fixed sample size across choices of $\beta_S$. The true NIE and NDE were fixed to $3$ and $0.5$, respectively. The simulation results are presented in Figure \ref{fig:mediation_comparison}. We see that with increasing selection bias strength, the NDE estimate produced by the naive approach is increasingly biased, whereas the selection adjusted estimator remains unbiased for the true effect, as expected. Both naive and selection adjusted estimates of the NIE are unbiased because in this DGP the NIE does not vary with $C$.

\section{Case Study: Disparities in Liver Transplant Decisions}

\subsection{Overview of the Transplantation Pipeline}

Liver transplantation is a life-saving procedure that requires patients with acute liver failure, cirrhosis, or liver cancer to make it through an arduous evaluation process with complicated criteria (e.g., laboratory testing, imaging, appointments). Patients that complete evaluation are reviewed by a transplant selection committee that discusses factors related to medical, surgical, and psychosocial risks to determine their eligibility for transplant listing. While many individuals are referred to transplant centers, only a fraction complete the evaluation process and are ultimately deemed eligible for a transplant. These are the ``listed'' patients who are candidates for transplant, anticipating to be matched with an appropriate donor organ if one becomes available. The process consists of four major stages. First, a patient is referred to a transplant clinic. Next, a workup is performed to evaluate the patient's candidacy based on considerations including disease condition, health biomarkers, and psychosocial evaluations. A committee meeting is held to discuss the candidate, and finally a listing decision is made. For more details on this process, consult \citet{strauss2022multicenter}.   

Patient dropout may occur between the start of the transplant referral stage and the final transplant listing decision. The likelihood of patient dropout in this process is nonrandom and evidence suggests that this may be linked to social determinants of health. Specifically, patients from underrepresented race/ethnicity groups that were socially disadvantaged -- as measured by lower neighborhood area deprivation index (ADI) scores -- were less likely to complete evaluation and experience a positive listing decision \citep{strauss2022multicenter,strauss2023critical}. These disparities are of great interest to the scientific community and are being investigated via novel approaches \citep{robitschek2024large}.

\subsection{Data and Research Questions}
We analyze data from a retrospective cohort of liver transplant candidates at Johns Hopkins Hospital. The target population consists of patients referred for liver transplantation from 1/1/2016-12/31/2017. The data includes patient information across five categories: baseline SDOH, SEP, disease-related variables, the outcome of psychosocial review evaluation, and final listing decision. Baseline SDOH variables include age, sex, race/ethnicity, and neighborhood ADI. Other social determinants such as educational attainment and native language were considered but ultimately not included in the analysis, since either reliable information was not recorded in available data or there was insufficient variation across these variables in the analysis. We use insurance status (private vs.\ not private) as a proxy for SEP, following previous research  \citep{robinson2014insurance,park2021sound}. 

The psychosocial assessment is captured by the social worker evaluation. Examples of factors they consider to make their assessment are substance use history, social support, psychiatric history, adherence to treatment plans, education, housing, and transportation. We summarize the outcome of this assessment with a binary flag indicating approval without any reservations versus concerns for transplant. 

Baseline characteristics of the full cohort, including a comparison between patients who completed the evaluation and those who dropped out, are presented in Table \ref{tab:cohort_characteristics}. For the cohort that completed the evaluation, we present characteristics stratified by the final listing decision in Table \ref{tab:evaluated_characteristics}.

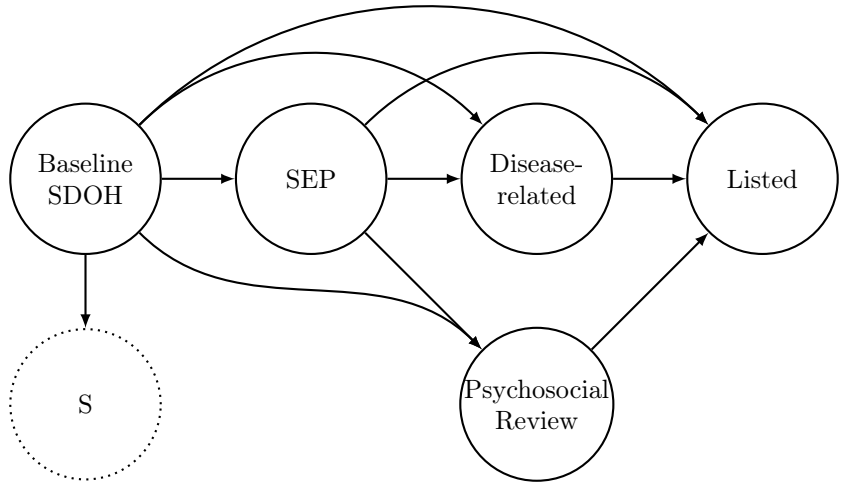
\begin{figure*}[h!]
        \centering
        \begin{tikzpicture}[->, >=latex, thick, node distance=2.5cm]
            \tikzset{node style/.style={circle, draw, minimum size=2cm, inner sep=0, font=\small, align=center}}
            
            \node (Baseline) [node style] at (0, 0) {Baseline\\ SDOH};
            \node (S) [node style, draw, dotted] at (0, -3) {S};  
            \node (SEP) [node style] at (3, 0) {SEP};
            \node (Disease) [node style] at (6, 0) {Disease-\\related};
            \node (Listed) [node style] at (9, 0) {Listed};
            \node (Psychosocial) [node style] at (6, -3) {Psychosocial\\ Review};
            
            \draw[->] (Baseline) -- (S);
            \draw[->] (Disease) -- (Listed);
            \draw[->] (Baseline) -- (SEP);
            \draw[->] (SEP) -- (Disease);
            
            \draw[->] (Baseline) to[out=45, in=135] (Listed);
            \draw[->] (SEP) to[out=45, in=135] (Listed);
            \draw[->] (Baseline) to[out=45, in=135] (Disease);
            
            \draw[->] (Baseline) to[out=315, in=135] (Psychosocial);
            \draw[->] (SEP) to[out=315, in=135] (Psychosocial);
            
            \draw[->] (Psychosocial) -- (Listed);
        \end{tikzpicture}
    \caption{Hypothesized relationships among liver transplant variables, visualized in a directed acyclic graph (DAG). SDOH = social determinants of health, SEP = socioeconomic position (here, insurance status).}
    \label{fig:measured-liver-transplant-graph}
\end{figure*}

Our focus is on the natural indirect and direct effects of SEP on transplant listing decision through the  psychosocial review outcome as a potential mediator. We assume the data-generating process may be approximately described by the DAG depicted in Figure \ref{fig:measured-liver-transplant-graph}. Though this model is certainly a simplification of a complex reality, we use it summarize some key variable relationships and highlight pathways of interest that may underlie socioeconomic disparities in listing status. The indirect effect of SEP through psychosocial review quantifies one potential mechanism by which SEP may affect listing decision. The ``direct'' effect parameter, in contrast, amalgamates the effect of SEP on listing decision through all other pathways (all pathways not through psychosocial review), which include disease-related pathways and potentially other mechanisms not mediated through any variables on the DAG.

Other pathways through which insurance may affect listing decision include disease etiology, disease severity, comorbidities, and surgical assessment since SEP affects which stage a patient is presented to the clinic. Since our effect is identified by the formula that only includes the pathway of interest, i.e. the pathway through psychosocial review, our effect estimation procedure does not require measurement of these disease-related variables.

We aim to estimate the target NIE and NDE parameters while accounting for any potential bias from sample selection. For patients who do not complete the evaluation process, there is no information on psychosocial review or listing outcomes, so our estimates of the NIE and NDE can only be based on the $361$ out of $497$ patients with complete data. We assume that selection into the analysis sample may be caused by some or all of the baseline SDOH as illustrated in the DAG. Fortunately, information on baseline SDOH is available for all patients referred to the transplantation clinic, so we may use this information to estimate selection weights and reweigh naive estimates of the NIE/NDE. Our DAG encodes the assumption that the available SDOH covariates satisfy the conditions of Theorem 2.

\begin{table*}[t]
\caption{Baseline Characteristics of the Study Cohort}
\label{tab:cohort_characteristics}
\tabcolsep=0pt
\begin{tabular*}{\textwidth}{@{\extracolsep{\fill}}lcccc}
\toprule
\textbf{Characteristic} & \textbf{Referred} & \textbf{Evaluated} & \textbf{Not Evaluated} & \textbf{p-value} \\
& \textbf{(n=593)} & \textbf{(n=419)} & \textbf{(n=174)} & \\
\midrule
Age, median [IQR] & 56 [49, 62] & 56 [49, 63] & 54 [48, 61] & 0.083 \\
Sex, Male, n (\%) & 350 (59.0) & 243 (58.0) & 107 (61.5) & 0.486 \\
National ADI, median [IQR] & 35 [21, 53] & 34 [19, 51.5] & 37 [25, 59.75] & 0.016 \\
Race/Ethnicity & & & & 0.012 \\
\quad White (non-Hispanic), n (\%) & 375 (63.2) & 280 (66.8)  & 95 (54.6) & \\
\quad Black (non-Hispanic), n (\%) & 122 (20.6) & 81 (19.3) & 41 (23.6) & \\
\quad Other, n (\%) & 96 (16.2) & 58 (13.8) & 38 (21.8) & \\
\bottomrule
\multicolumn{5}{l}{\footnotesize{IQR: interquartile range; SD: Standard Deviation; ADI: Area Deprivation Index.}} \\
\multicolumn{5}{l}{\footnotesize{P-values compare the `Evaluated' and `Not Evaluated' groups.}} \\
\end{tabular*}
\end{table*}

\begin{table*}[t]
\caption{Characteristics of the Evaluated Cohort, Stratified by Listing Decision}
\label{tab:evaluated_characteristics}
\tabcolsep=0pt
\begin{tabular*}{\textwidth}{@{\extracolsep{\fill}}lcccc}
\toprule
\textbf{Characteristic} & \textbf{Evaluated} & \textbf{Not Listed} & \textbf{Listed} & \textbf{p-value} \\
& \textbf{(n=419)} & \textbf{(n=172)} & \textbf{(n=247)} & \\
\midrule
Age, median [IQR] & 56 [49, 63] & 57 [50, 63] & 56 [49, 62.5] & 0.245 \\
Sex, Male, n (\%) & 243 (58.0) & 100 (58.1)  & 143 (57.9) & 1.000 \\
National ADI, median [IQR] (SD) & 34 [19, 51.5] & 38 [20.75, 55] & 30 [19, 50] & 0.027 \\
Private Insurance, n (\%) & 171 (40.8) & 48 (27.9) & 123 (49.8) & $<$0.001 \\
Psychosocial Assessment, mean (SD) & 0.58 (0.49) & 0.39 (0.49) & 0.70 (0.46) & $<$0.001 \\
Race/Ethnicity & & & & 0.006 \\
\quad White (non-Hispanic), n (\%) & 280 (66.8) & 101 (58.7) & 179 ( 72.5) & \\
\quad Black (non-Hispanic), n (\%) & 81 (19.3) & 45 (26.2) & 36 ( 14.6) & \\
\quad Hispanic/Other, n (\%) & 58 (13.8) & 26 (15.1) & 32 ( 13.0) & \\
\bottomrule
\multicolumn{5}{l}{\footnotesize{IQR: interquartile range; SD: Standard Deviation; ADI: Area Deprivation Index.}} \\
\multicolumn{5}{l}{\footnotesize{P-values compare the `Not Listed' and `Listed' groups.}} \\
\end{tabular*}
\end{table*}


\begin{figure*}[h!]
    \centering

    \begin{tikzpicture}[->, >=latex, thick, node distance=3cm]
        \tikzset{node style/.style={circle, draw, minimum size=2cm, inner sep=0pt, font=\small, align=center}}
        \tikzset{red node style/.style={circle, draw=red, minimum size=2cm, inner sep=0pt, font=\small, align=center}}
    
        \node (Baseline) [node style] at (0, 0) {Baseline\\ SDOH};
        \node (S) [node style, draw, dotted] at (0, -3) {S};  
        \node (SEP) [node style] at (4, 0) {SEP};
        \node (Disease) [node style] at (8, 0) {Disease-\\related};
        \node (Listed) [node style] at (12, 0) {Listed};
        \node (Psychosocial) [node style] at (8, -3) {Psychosocial\\ Review};
        \node (Transportation) [red node style] at (4, -3) {Transportation};
        \node (HealthLit) [red node style] at (4, -6) {Health\\ Literacy};
    
        \draw[->] (Disease) -- (Listed);
        \draw[->] (Baseline) -- (SEP);
        \draw[->] (Baseline) -- (S);
        \draw[->] (SEP) -- (Disease);
    
        \draw[->] (Baseline) to[out=45, in=135] (Listed);
        \draw[->] (SEP) to[out=45, in=135] (Listed);
        \draw[->] (Baseline) to[out=45, in=135] (Disease);
    
        \draw[->] (Baseline) to[out=315, in=135] (Psychosocial);
        \draw[->] (SEP) to[out=315, in=135] (Psychosocial);
    
        \draw[->] (Psychosocial) to[out=0, in=270] (Listed);

        \draw[->] (Baseline) to[out=315, in=135] (Transportation);
        \draw[->] (SEP) to[out=215, in=135] (Transportation);
        \draw[->] (Transportation) to[out=45, in=215] (Disease);
        \draw[->] (Transportation) to[out=0, in=180] (Psychosocial);
    
        \draw[->] (Baseline) to[out=315, in=135] (HealthLit);
        \draw[->] (SEP) to[out=215, in=135] (HealthLit);
        \draw[->] (HealthLit) to[out=45, in=215] (Disease);
        \draw[->] (HealthLit) to[out=0, in=180] (Psychosocial);
    
        \draw[->] (HealthLit) to[out=0, in=270] (Listed);
    \end{tikzpicture}

    \caption{An expanded hypothetical DAG for the liver transplantation variables.}
    \label{fig:total-liver-transplant-graph}
\end{figure*}
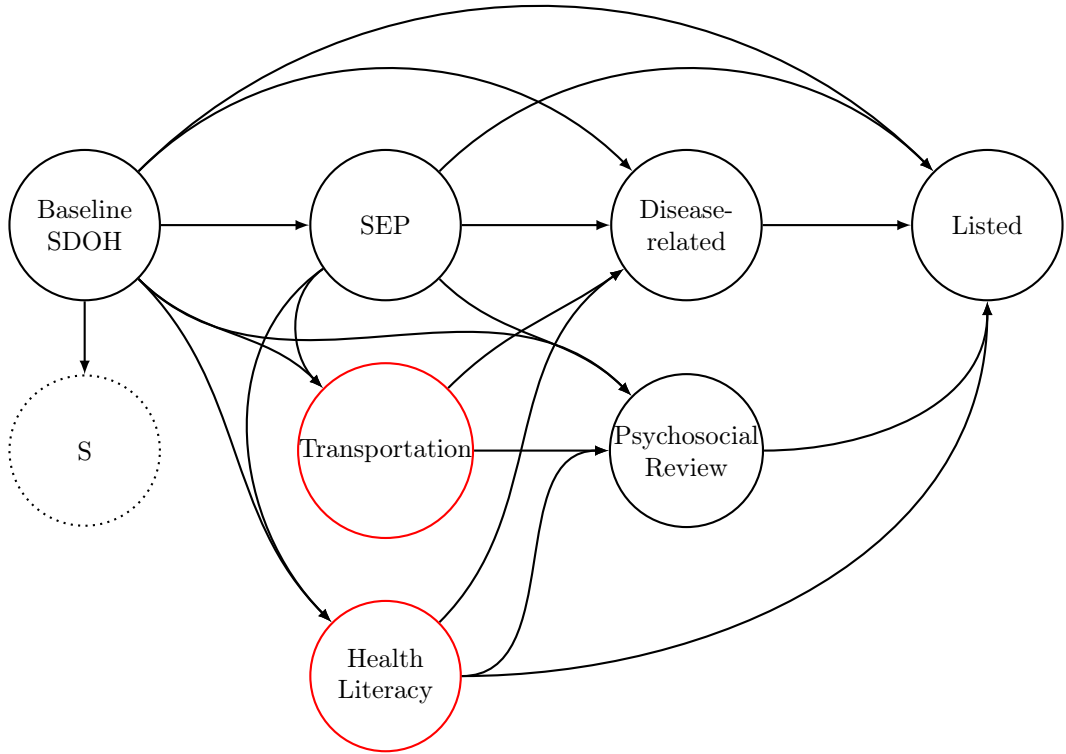

\subsection{Analysis}



Certain categorical variables were consolidated due to sample size limitations. Psychosocial review is assessed on a four point scale on a dimension that measures social worker support of the application, but this was binarized to indicate social worker support of the application versus caution. Race/ethnicity categories were reduced to three: White (non-Hispanic), Black (non-Hispanic), and Hispanic or Other.

All estimates were obtained using a modification of the counterfactual imputation estimator in the R package \texttt{CMAverse} \citep{shi2021cmaverse} that has been modified to allow incorporation of selection weights. This estimator requires the specification of two nuisance models: a model for the mediator and a model for the outcome, both conditional on all preceding variables in the DAG. We use logistic regression models with pairwise interactions for both. Selection weights are also estimated using logistic regression with pairwise interactions among SDOH variables. Results obtained using our selection bias adjustment formula alongside a naive estimation of the NIE/NDE are displayed in Table \ref{tab:effects_table}. We report effect estimates on the risk ratio scale.

\begin{table}[t]
\caption{Summary of mediation effects of insurance status on transplant listing, comparing standard estimates to estimates adjusted for selection bias.}
\label{tab:effects_table}
\centering
\begin{tabular*}{0.8\textwidth}{@{\extracolsep{\fill}}lc}
\toprule
\textbf{Effect} & \textbf{Risk Ratio (95\% CI)} \\
\midrule
\multicolumn{2}{l}{\textit{Standard (Naive) Mediation Analysis}} \\
Natural Indirect Effect (NIE-RR) & 1.11 (1.04, 1.19) \\
Natural Direct Effect (NDE-RR) & 1.26 (1.06, 1.56) \\
Total Effect & 1.39 (1.20, 1.63) \\
\midrule
\multicolumn{2}{l}{\textit{Selection Adjusted Mediation Analysis}} \\
Natural Indirect Effect (NIE-RR) & 1.10 (1.03, 1.19) \\
Natural Direct Effect (NDE-RR) & 1.26 (1.05, 1.49) \\
Total Effect & 1.39 (1.19, 1.67) \\
\bottomrule
\multicolumn{2}{p{0.8\textwidth}}{
  \begin{minipage}[t]{\linewidth}
    \footnotesize
    CI: Confidence Interval. Reported intervals are at the 95\% threshold. \\
    The NIE-RR represents the effect of insurance status mediated through the psychosocial assessment. \\
    The NDE-RR represents the effect of insurance status through all other pathways.
  \end{minipage}
} \\
\end{tabular*}
\end{table}

The results indicate that there is a significant effect of insurance status on listing decision. In the naive analysis, the estimated risk ratio for the NIE is $1.11 \; (1.04, 1.19)$ and for the NDE is $1.26 \; (1.06, 1.56)$. Thus, given our model assumptions, the evidence supports the existence of both indirect and direct effects of insurance status on listing decisions with respect to psychosocial review. The effect estimates remain largely unchanged after adjusting for selection bias using our estimated weights: the selection adjusted NIE and NDE estimates are $1.10 \; (1.03, 1.19)$ and $1.26 \; (1.05, 1.49)$, respectively. This suggests sample selection has minimal on the effect estimates in this data and the NDE and NIE effect sizes are substantial even taking sample selection into account.

One important limitation in our study is that our model assumes there are no post-exposure variables that may act as recanting witnesses. It is possible that some additional factors, including a patient's access to transportation and their health literacy, may be affected by SEP and also affect the outcomes of psychosocial review. This possibility is illustrated in Figure \ref{fig:total-liver-transplant-graph}. Unfortunately, data on these factors is not available in our study and so we cannot assess the potential impact of these confounders on our estimates. In future work, data collection efforts may focus on extracting information on some of these more challenging-to-measure variables that are post-exposure. 

\section{Conclusion}

This work introduces selection adjusted formulae for mediation and path-specific effects and establishes a set of sufficient conditions for the validity of these formulae by extending previous work on adjustment for selection bias. There may however be other conditions under which the target effects are identified in the presence of selection bias. Future work may focus on identification strategies that are more complicated than adjustment but known to have completeness guarantees \citep{shpitser2016causal,shpitser2018identification,correa2019identification}. Compete algorithms for studying mediation analysis in the presence of selection bias is a topic left to future work.

Our application to a liver transplant case study revealed that that socioeconomic position (as captured by insurance status) has potentially strong indirect effects on listing decision through the outcome of psychosocial review and that selection bias did not have a substantial impact on these results. These findings are consistent with existing work which show insurance status as a predictor for liver transplant listing \citep{stepanova2020outcomes,robitschek2024large}. Furthermore, insurance status is linked to a variety of socioeconomic characteristics including income level \citep{yilma2023community}, housing status \citep{flanary2025access}, social support \citep{bangarusocial}, and neighborhood poverty levels \citep{flanary2025access}. Although the psychosocial review considers these factors, our data did not parse out which of these play a significant role. Future research to collect this data and better understand which specific aspects are most impactful in the psychosocial review can guide targeted clinical interventions. This also highlights the potential areas where latent factors may be at play, such as racial implicit biases, with patients who may have similar socioeconomic backgrounds receiving differential priority for listing \citep{flanary2025access}. These findings highlight the need to further explore strategies for inequity mitigation in liver transplant listing.


\begin{appendices}

\section{Appendix}

\subsection{Lemmas}

In this subsection we prove two lemmas. We being by describing the relationship between the second condition in Theorems 2 and 3 and the prohibition of recanting districts \citep{shpitser2013counterfactual}. A district in graph $G$ is a set of vertices that are all bidirected-connected in $G$.

\begin{definition}
\textbf{Recanting District}. Given a causal graph $G$, with a disjoint set of vertices $X, Y \subseteq V$ and a set of proper causal paths $\pi$, the district $D$ in $G$ is said to be a recanting district for identifying the $\pi$-specific effect of $X$ on $Y$ if there exist $D_i, D_j \in D$ (possibly $D_i=D_j$), $X_i \in X$, and $Y_i, Y_j \in Y$ (possibly $Y_i=Y_j$) such that there is a proper causal path $X_i \rightarrow D_i \rightarrow ... \rightarrow Y_i$ in $\pi$ and a proper causal path $X_i \rightarrow D_j \rightarrow ... \rightarrow Y_j$ not in $\pi$.
\end{definition}

In the typical setting with no selection bias, the $\pi$-specific effect of $X$ on $Y$ is expressible as a functional of interventional densities if and only if a recanting district is not present in $G$ \citep{shpitser2013counterfactual}. Our Theorems 2 and 3 impose a stricter assumption in the setting with selection bias. This assumption requires that backdoor paths between mediators and outcome are blocked by $\{Z, S\}$. This is closely related to the ``no recanting districts'' condition since the existence a recanting district would imply that there is an unblocked backdoor path between a mediator and outcome. 

\begin{lemma}
Given a causal graph $G$ augmented with selection node $S$, if there exists a recanting district $D$ for identifying the $\pi$-specific effect of $X$ on $Y$, then for some $M$ on a causal path from $X$ to $Y$, $Y$ is m-connected to $M$ given $\{Z,S\}$ in $G^{pbd}_{(X,M),Y}$.
\end{lemma}

\textbf{Proof: }
Suppose that there exists a recanting district $D$ for the $\pi$-specific effect of $X$ on $Y$ in $G$. By definition, there exist $D_i,D_j \in D$, such that
\[
\begin{aligned}
X_i &\rightarrow D_i \rightarrow \pi_i \rightarrow Y_i \\
X_i &\rightarrow D_j \rightarrow \pi_j \rightarrow Y_j
\end{aligned} \]
Where $\pi_i, \pi_j$, $\pi_i \not = \pi_j$ denote directed paths from which $D_i, D_j$ can reach $Y_i, Y_j$, respectively. As $D_i, D_j \in D$ are in a district, they are connected by a sequence of bidirected edges, either directly or via some intermediate nodes. Define $\pi_d$ to be the bidirected path connecting $D_i, D_j$ such that $D_i \leftrightarrow \pi_d \leftrightarrow D_j$. Then in $G$ there exists an m-connecting backdoor path from some vertex on $\pi_i$ to $Y_j$:  
\[
\begin{aligned}
\pi_i \leftarrow D_i \leftrightarrow \pi_d \leftrightarrow D_j \rightarrow \pi_j \rightarrow y_j
\end{aligned}
\]
This backdoor path cannot be blocked by $\{Z,S\}$ since $\{Z, S\}$ cannot contain descendants of mediators and if any elements of $\{Z,S\}$ are on $\pi_d$ then they are colliders on the path. In the case that $D_i = D_j$, then the relevant backdoor path is $\pi_i \leftarrow D_i \rightarrow \pi_j \rightarrow y_j$. Let $M$ be a vertex on $\pi_i$. This m-connecting path is also in $G^{pbd}_{(X,M),Y}$, so $M$ is m-connected to $Y$ given $\{Z,S\}$.

\begin{lemma}
\label{prop:med-exposure-outcome-indep}


If $(X \perp_m Y \mid Z, S)_{G^{pbd}_{X, Y}}$ and $(M_i \perp_m Y \mid Z, S)_{G^{pbd}_{(X, M_i), Y}}$, then the following m-separations are implied:

\begin{enumerate}[(a)\hspace{0.5em}]
    \item $(X \perp_m Y \mid Z, S)_{G^{e, pbd}_{X, Y}}$
    \item $(X^e \perp_m Y \mid Z, S)_{G^{e, pbd}_{X^e, Y}}$
    \item $(M_i \perp_m Y \mid Z, S)_{G^{e, pbd}_{(X^e, M_i), Y}}$
    \item $(M_i \perp_m Y \mid Z, S)_{G^{e, pbd}_{(X, M_i), Y}}$
    \item $(X^e_{m_i} \perp_m Y \mid X^e_y, M_i, Z, S )_{G^{e, pbd}_{\overline{X^e_{y}}, X^e_{m_i}, Y}}$
    \item $(M_i \perp_m Y \mid \Pa^{M}(M_i), Z, S)_{G^{e, pbd}_{(X^e, M_i), Y}}$
    \item $(M_i \perp_m X^e \mid Z, S)_{G^{e, pbd}_{X^e, M_i}}$
    \item $(M_i \perp_m X^e \mid \Pa^{M}(M_i), Z, S)_{G^{e, pbd}_{X^e, M_i}}$
\end{enumerate}
\end{lemma}

\textbf{Proof: }
\begin{enumerate}[(a)\hspace{0.5em}]
    \item If there were an m-connecting path to violate $(X \perp_m Y \mid Z, S)_{G^{e, pbd}_{X, Y}}$, then that path would also violate $(X \perp_m Y \mid Z, S)_{G^{pbd}_{X, Y}}$, since the two graphs only by additional paths through extended extended nodes between $X$ and $M, Y$, which do not affect the m-connecting status of a path given $\{Z,S\}$.
    \item Any path that would m-connect $X^e$ and $Y$ in $G^{pbd}_{X^e, Y}$ would similarly m-connect $X$ and $Y$ in $G^{pbd}_{X, Y}$ by beginning with an edge into $X$ in lieu of $X \rightarrow X^e$, which is the only possible edge into $X^e$ in $G^{pbd}_{X^e, Y}$.
    \item Note that $G^{pbd}_{(X, M_i), Y}$ and $G^{e, pbd}_{(X^e, M_i), Y}$ are equivalent with the exception of the additional extended nodes $X^e$ and additional edges $X \rightarrow X^e$. Edges to $X^e$ do not change the collider/non-collider status of other nodes in the graph. Any path that would m-connect $M_i$ and $Y$ in $G^{e, pbd}_{(X^e, M_i), Y}$ would similarly m-connect $M_i$ and $Y$ in $G^{pbd}_{(X, M_i), Y}$.
    \item $(M_i \perp_m Y \mid Z, S)_{G^{e, pbd}_{(X, M_i), Y}}$ removes edges into $X^e$ whereas $G^{e, pbd}_{(X^e, M_i), Y}$ in case (c) removes edges out of $X^e$. $X^e$ are intermediary extended nodes that intercept paths from $X$ to $M_i$ and $Y$. Removing the edges out of or into $X^e$ does not change whether $M_i$ and $Y$ are m-connected since the only possibly relevant paths are backdoor paths into $M_i$ not through $X$.
    \item $G^{e}_{\overline{X^e_{y}}}$ is the graph removes the edge into $X^e_{y}$. $G^{e, pbd}_{\overline{X^e_{y}}, X^e_{m_i}, Y}$ additionally removes the edge out of $X^e_{m_i}$.
    $G^{e, pbd}_{X^e, Y}$, in which the m-separation between $X^e$ and $Y$ is satisfied by (b), removes edges out of $X^e_{m_i}$ and $X^e_y$. $X^e_y$ only has an edge to $Y$ in $G^{e, pbd}_{\overline{X^e_{y}}, X^e_{m_i}, Y}$ and is a leaf node in $G^{e, pbd}_{X^e, Y}$, therefore the edge differences w.r.t $X^e_y$ between these graphs do not affect collider/non-collider status of m-connecting paths between $X^e_{m_i}$ and $Y$. The only edge connecting to $X^e_{m_i}$ in both of these graphs is $X \rightarrow X^e_{m_i}$, so an m-connecting path between $Y$ and $X$ would imply an m-connecting path between $Y$ and $X^e_{m_i}$ in these graphs. Case (a) implies $(X \perp_m Y \mid Z, S)_{G^{e, pbd}_{X^e, Y}}$ therefore $(X^e_{m_i} \perp_m Y \mid X^e_y, M, Z, S)_{G^{e, pbd}_{\overline{X^e_{y}}, X^e_{m_i}, Y}}$ provided that introducing $M_i$ to the conditioning set does not open any paths that would otherwise be blocked. However, $M_i$ cannot act as a collider and open any paths as this would contradict (c). Therefore $X^e_{m_i}$ and $Y$ must be m-separated in $G^{e, pbd}_{\overline{X^e_{y}}, X^e_{m_i}, Y}$ by $X^e_y, M_i, Z, S$.
    \item Given (c), an m-connecting path can only induced if the new conditioning set $\Pa^{M}(M_i), Z, S$ includes a collider such that a path from $M_i$ to $Y$ is now open. The candidate m-connecting path must connect $M_i$ and $Y$, traversing some $M_k \in \Pa^{M}(M_i)$, which is a collider on the path. $Z$ and $S$ cannot be non-colliders on the path or else it would be blocked. The subpath from $M_k$ to $Y$ would contradict $(M_k \perp_m Y \mid Z, S)_{G^{e,pbd}_{(X^e,M_k),Y}}$.
    \item $G^{e, pbd}_{X^e, M_i}$ removes edges out of extended nodes that are along proper causal paths to $M_i$. Call this relevant set of extended nodes $X^e_{\An(M_i)} = \{ X^e \mid X^e \in \An(M_i) \}$. Because $X^e$ only has edges from $X$ and to its children, it suffices to prove m-separation between (1) $M_i$ and $X$ and between (2) $M_i$ and the set of extended nodes that are not ancestors of $M_i$, as all other extended nodes are only connected to $X$. Regarding (1), any candidate m-connecting path from $X$ to $M_i$ must be noncausal as this is a proper backdoor graph, $G^{e, pbd}_{X^e, M_i}$. If this candidate path has a subpath through $X^e_{-\An(M_i)} = \{X^e \mid X^e \not \in \An(M_i) \}$, we move to case (2). If not, then this path exists in $G^{e, pbd}_{X^e, Y}$ and concatenating the path between $X$ and $M_i$ with the directed path from $M_i$ to $Y$ violates (b). Regarding (2), if this candidate path m-connects $X^e_{-\An(M_i)}$ and $M_i$, it must be via edges out of $X^e_{-\An(M_i)} \rightarrow M_k \dots M_i$ where $M_k \in \Ch(X^e_{-\An(M_i)})$ (since the paths through $X$ are already addressed). This path cannot be causal as $X^e_{-\An(M_i)}$ is the set of extended nodes non-ancestral of $M_i$. If a noncausal path connects $M_i$ and some $M_k \in \Ch(X^e_{-\An(M_i)})$, then this path would violate (c) for $M_i$ via the noncausal path from $M_i$ to $M_k$ concatenated with the path $M_k$ to $Y$. 
    \item Given (g), an m-connecting path can only be induced if the new conditioning set $\Pa^{M}(M_i), Z, S$ includes a collider such that the path from $X^e$ to $M_i$ is now open. The candidate m-connecting path must connect $M_i$ and $X^e$, traversing some $M_k \in \Pa^{M}(M_i)$, which is a collider on the path. $Z$ and $S$ cannot be non-colliders on the path or else it would be blocked. We frame our argument similar to (g): we consider candidate m-connecting paths (1) between $X$ and $M_i$ and (2) $M_i$ and the set of extended nodes that are not ancestors of $M_i$, denoted $X^e_{-\An(M_i)}$. Case (1): if the subpath $M_k$ to $X^e$ does not intersect $X^e_{-\An(M_i)}$, then this subpath contradicts $(M_k \perp_m X^e \mid Z, S)_{G^{e, pbd}_{X^e, M_k}}$. Case (2): see (g) case (2). No such m-connecting path can exist.
\end{enumerate}

\subsection{Proof of Theorem \ref{theorem:GACiff}}

First, the forwards direction: for the following three GAC conditions, we assume they hold in $G$ and prove they must hold in $G^e$.

{By assumption, no element in $Z$ is a descendant in $G_{\overline{X}}$ of any $W \notin X$ lying on a proper causal path from $X$ to $Y$.
} Suppose there exists a $Z$ that is a descendant of some $W$ on a proper causal path from $X^e$ to $Y$ in $G^e_{\overline{X^e}}$. If $Z$ is a descendant of some $W$ lying on a proper causal path from $X^e$ to $Y$, then $Z$ is also a descendant of a $W$ on a proper causal path from $X$ to $Y$, since $X$ is a parent of $X^e$ and they share all descendants. Contradiction.

{By assumption, all non-causal paths in $G$ from $X$ to $Y$ are blocked by $Z$ and $S$.} If all non-causal paths 
$$
X \leftarrow ... \rightarrow Y
$$
in G are blocked by $Z, S$, then the same $Z, S$ will block all non-causal paths from 
$$
X^e \leftarrow X \leftarrow ... \rightarrow Y
$$
in $G^e$ as $Z$ and $S$ would have the same non-collider or non-collider status on the concatenated path as they did in the original path.

{By assumption, $Z^T$ m-separates $Y$ from $S$ in the proper backdoor graph, i.e $(Y \perp S \mid Z^T)_{G^{pbd}_{X, Y}}$.}
Let us first note the difference between $G^{pbd}_{X, Y}$ and $G^{e, pbd}_{X^e, Y}$. The extended graph includes $X^e$ and the edges from $X$ to $X^e$ to children (in $G$) of $X$, however the proper backdoor graph removes the first edge from every proper causal path of a given starting set and ending set. In the first graph, this removes the first edge in any proper causal path of the form 
$$
X \rightarrow M \rightarrow ... \rightarrow Y \text{ or } X \rightarrow Y
$$
In the extended graph, we remove the second edge (as the starting point for the proper backdoor construction is the set $X^e$) in any path of the form 
$$
X \rightarrow X^e_m \rightarrow M \rightarrow ... \rightarrow Y \text{ or } X \rightarrow X^e_y \rightarrow Y
$$
Thus, the only difference between the two graphs is that $G^{e, pbd}_{X^e, Y}$ includes edges of the form 
$X \rightarrow X^e$. Just as above, $Z^T$ has the same non-collider or non-collider status along all paths from $Y$ to $S$ in $G^e$ as it did in $G$, therefore it will m-separate $Y$ and $S$ in the extended proper backdoor graph.

Next, the backwards direction: here we show if the GAC hold in $G^e$ they must hold in $G$.

{By assumption, no element in $Z$ is a descendant in $G^e_{\overline{X^e}}$ of any $W \notin X^e$ lying on a proper causal path from $X^e$ to $Y$.} Suppose there exists a $Z$ that is a descendant of some $W$ on a proper causal path from $X$ to $Y$. If $Z$ is a descendant of some $W$ lying on a proper causal path from $X$ to $Y$, then then we have that there is a $W$ along the path $X^e$ to $Y$ below. 
$$
X \rightarrow X^e \rightarrow ...  \rightarrow Y
$$
$W$ lies on a proper causal path from $X^e$ to $Y$ in $G^e$ also, and so $Z$ is a descendant of such a $W$ in $G^e$. Contradiction.

{By assumption, all non-causal paths in $G$ from $X^e$ to $Y$ are blocked by $Z$ and $S$.
} If all non-causal paths are blocked from
$$
X^e \leftarrow X \leftarrow ... \rightarrow Y
$$
by $Z, S$, then it is evident that the same $Z, S$ will block all non-causal paths from 
$$
X \leftarrow ... \rightarrow Y
$$
as $Z, S$ would have the same collider or non-collider status on the original path as they did in the concatenated path (as $X^e$ does not belong to $Z, S$ and is not related to them outside of through its only parent $X$); therefore they will block all non-causal paths in the original path. 

{By assumption, $Z^T$ m-separates $Y$ from $S$ in the proper backdoor graph, i.e $(Y \perp S \mid Z^T)_{G^{e, pbd}_{X^e, Y}}$.
} The only difference between the two graphs is that $G^{e, pbd}_{X_e, Y}$ includes edges of the form $ X \rightarrow X^e $ as discussed above. $Z^T$ has the same non-collider or non-collider status along the relevant paths in both $G^{e, pbd}_{X_e, Y}$ and $G^{ pbd}_{X, Y}$, so the conclusion follows.

\subsection{Proof of Theorem \ref{theorem:mediationadmg}}

Let us rewrite the distribution of nested counterfactual $Y(x, M(x'))$ in do-notation using two extended nodes: $X_y^e$ and $X_m^e$. These extended nodes intercept two paths, $X \rightarrow Y$ and $X \rightarrow M$, such that these paths become $X \rightarrow X_y^e \rightarrow Y$ and $X \rightarrow X^e_m \rightarrow M$, respectively, in $G^e$. We have the equivalence: 

$$
E[Y(x, M(x')] = E[ Y \mid \doo(X^e_y = x, X^e_m = x')] 
$$

Using Theorem \ref{theorem:GACiff}, if and only if the GAC conditions hold for $X^e_y, X^e_m$, this quantity is equal to: 

$$
E[Y \mid \doo(X^e_y = x, X^e_m = x')] = \sum_z E[Y \mid \doo(X^e_y = x, X^e_m = x'), z, S=1] p( z \setminus z^T \mid z^T, S=1) p(z^T)
$$

We introduce an additional marginalization over $M$ and then simplify the expression using the do-calculus.

\begin{align*}
    &p(y \mid \doo(X^e_y = x, X^e_m = x'))\\ 
&= \sum_{z, m} p(y \mid \doo(X^e_y = x, X^e_m = x'), m, z, S=1)p(m \mid \doo(X^e_y=x, X^e_m=x'), z, S=1)  p( z \setminus z^T \mid z^T, S=1) p(z^T) \\
&=^{(1)} \sum_{z, m} p(y \mid \doo(X^e_y = x), m, z, S=1)  p(m \mid \doo(X^e_y = x, X^e_m = x'), z, S=1)  p( z \setminus z^T \mid z^T, S=1) p(z^T)\\
&=^{(2)} \sum_{z, m} p(y \mid \doo(X^e_y = x), m, z, S=1)   p(m \mid \doo(X^e_m = x'), z, S=1)  p( z \setminus z^T \mid z^T, S=1) p(z^T)\\
&=^{(3)} \sum_{z, m} p(y \mid X^e_y = x, m, z, S=1)   p(m \mid X^e_m = x', z, S=1)  p( z \setminus z^T \mid z^T, S=1) p(z^T)\\
&= \sum_{z, m} p(y \mid X = x, m, z, S=1)   p(m \mid X = x', z, S=1)  p( z \setminus z^T \mid z^T, S=1) p(z^T)
\end{align*}

The equality $=^{(1)}$ follows from applying Rule 3 of the do-calculus to remove the $\doo(X^e_m)$ in the $p(y \mid \doo(X^e_y, X^e_m), m, z, S=1)$ term. Define $G^e_{\overline{X_y^e}}$ as the modified copy of $G^e$ where the edge into $X_y^e$ is removed. (The edge into $X^e_m$ is not removed, as it is an ancestor of $M$ which belongs to the conditioning set.) For Rule 3 to hold, $Y$ and $X^e_m$ must be m-separated in $G^e_{\overline{X_y^e}}$ by $\{ X^e_y, M, Z, S \}$. Non-causal paths from $X^e_m$ to $Y$ are m-separated in $G^{e, pbd}_{\overline{X_y^e}, X^e_m, Y}$ given $\{X^e_y, M, Z, S\}$ by Lemma \ref{prop:med-exposure-outcome-indep}(e). The only possibly m-connecting path between $X^e_m$ and $Y$ is the causal path $X^e_m \rightarrow M \rightarrow Y$. The m-separation holds due to conditioning on $M$. Thus, the condition for Rule 3 applies.

The equality $=^{(2)}$ follows using Rule 3 to remove $\doo(X^e_y = x)$ in the $p(m \mid \doo(X^e_y, X^e_m), z, S=1)$ term. Define $G^e_{\overline{X_m^e},\overline{X_y^e}}$ as the graph that removes edges into $X^e_m$ and $X^e_y$. $M \perp X^e_y \mid Z,S$ holds in this graph. We have by Lemma \ref{prop:med-exposure-outcome-indep}(d) that there is no backdoor m-connecting path between $M$ and $Y$ given $Z, S$ in $(M \perp_m Y \mid Z, S)_{G^{e, pbd}_{(X, M), Y}}$. $G^{e, pbd}_{(X, M), Y}$ differs from $G^e_{\overline{X_m^e},\overline{X_y^e}}$ in that the latter also includes the edge $M \rightarrow Y$. However, this additional edge does not m-connect $X^e_y$ and $M$ since the path $M \rightarrow Y \leftarrow X^e_y$ is blocked by the collider $Y$. Thus, the condition for Rule 3 applies.

The equality $=^{(3)}$ follows by Rule 2. In order to use Rule 2 on the term $p(y \mid \doo(X^e_y), m, z, S=1)$, we must have $Y \perp X^e_y \mid M, Z, S $ hold in $G^e_{\underline{X^e_y}}$, the graph with edges out of $X^e_y$ removed. Removing the edge out of $X^e_y$ leaves no way for $Y$ to be m-connected to the extended node as $\{ Z, S \}$ block all back door paths between $Y$ and $X$ (the only other node connected to $X^e_y$) by Lemma \ref{prop:med-exposure-outcome-indep}(a). 

For $p(m \mid \doo(X^e_m), z, S=1)$, we must have $M \perp X^e_m \mid Z, S$ hold in $G^e_{\underline{X^e_m}}$, the graph with edges out of $X^e_m$ removed. No causal paths from $X^e_{m}$ to $M$ exist in this modified graph due to the edge removal. No noncausal m-connecting path exists between $X^e_{m}$ and $M$ in $G^e_{\underline{X^e_m}}$ when conditioning on $Z, S$. $G^{e}_{\underline{X^e_{m_i}}}$ is the graph that removes the edge out of $X^e_{m_i}$. $G^{e}_{\underline{X^e_{m_i}}}$ is identical to $G^{e, pbd}_{\underline{X^e_{m_i}}, X^e_{m_i}, M_i}$. This differs from $G^{pbd}_{X^e, Y}$ since the only edge that is removed is the one out of $X^e_{m_i}$ rather than out of all $X^e$. $X^e_{m_i}$ only has an edge to $X$; any m-connecting path between $M_i$ and $X^e_{m_i}$ must include $X \rightarrow X^e_{m_i}$. It suffices to show that $M_i$ and $X$ are not m-connected in $G^{e}_{\underline{X^e_{m_i}}}$ to prove the m-separation between $M_i$ and $X^e_{m_i}$. Say there were an m-connecting path from $M_i$ to $X$ given $\{Z,S\}$ in $G^{e}_{\underline{X^e_{m_i}}}$. This path cannot begin with $M_i \rightarrow Y$ since $Y$ would be a collider on that path. Concatenating that path with $M_i \rightarrow Y$ would lead to an m-connection from $Y$ to $X$ and violate Lemma \ref{prop:med-exposure-outcome-indep} (b). Note that (b) refers to $G^{e,pbd}_{X^e,Y}$ rather than $G^{e}_{\underline{X^e_{m_i}}}$ but these only differ by the $X^e_y \rightarrow Y$ edge which cannot affect the m-connection status of this path.

The final equality holds by the definition of the extended model with $X^e_y$ and $X^e_m$.

\subsection{Proof of Theorem \ref{theorem:pseadjustment}}

Let us rewrite the distribution of the counterfactual $Y(\pi, x, x')$ in do-notation using a vector of extended nodes: $X^e$. Here, $X^e = \{ X_y^e, X_m^e\}$ where $X_y^e$ is a single extended node, $X_m^e$ is a vector of extended nodes (one for each mediator on a proper causal path from $X$ to $Y$),
and $\pi$ is a set of paths that will receive the treatment $x$ while paths not in $\pi$ will take on treatment $x'$. $M$ is the set of mediators of size $|M|$. These extended nodes intercept paths of the form $X \rightarrow Y$ and $X \rightarrow M_i$ for some $M_i \in M$, such that these paths become $X \rightarrow X_y^e \rightarrow Y$ and $X \rightarrow X^e_{m_i} \rightarrow M_i$, respectively, in $G^e$. Denote $x^{\pi}$ as the vector of interventional values being applied to each extended node in $X^e$ depending on if that particular $X \rightarrow X^e$ is in $\pi$ or not.

$$
p(Y(\pi, x, x')) = p(y \mid \doo(X^e = x^{\pi})) 
$$

Using Theorem \ref{theorem:GACiff}, if and only if the GAC conditions hold for $X^e_y, X^e_m$, this quantity is equal to: 

\begin{align*}
p(y \mid \doo(X^e = x^{\pi})) =  \sum_z p(y \mid \doo(X^e = x^{\pi}), z, S=1) \; p( z \setminus z^T \mid z^T, S=1) \; p(z^T) 
\end{align*}

Assume that $M$ is topologically ordered such that $M_i$ cannot be a parent of some $M_j$ if $i > j$. We now introduce an additional marginalization over $M$ and then simplify the expression using the do-calculus.

\begin{align*}
    &p(y \mid \doo(X^e = x^{\pi}))\\ 
&= \sum_{z, m} p(y \mid \doo(X^e = x^{\pi}), m, z, S=1)p(m \mid \doo(X^e = x^{\pi}), z, S=1)  p( z \setminus z^T \mid z^T, S=1) p(z^T) \\
&=^{(1)} \sum_{z, m} p(y \mid \doo(X^e = x^{\pi}), m, z, S=1) \left[ \prod_{i=1}^{ |M| } p(m_i \mid \doo(X^e = x^{\pi}), m_1, ..., m_{i-1}, z, S=1) \right] \\ 
& \quad \quad p( z \setminus z^T \mid z^T, S=1) p(z^T) \\
&=^{(2)} \sum_{z, m} p(y \mid \doo(X^e = x^{\pi}), m, z, S=1) \left[ \prod_{i=1}^{ |M| } p(m_i \mid \doo(X^e = x^{\pi}), \Pa^{M}(M_i), z, S=1) \right] \\  
& \quad \quad p( z \setminus z^T \mid z^T, S=1) p(z^T) \\
&=^{(3)} \sum_{z, m} p(y \mid \doo(X^e = x^{\pi}), m, z, S=1) \left[ \prod_{i=1}^{ |M| } p(m_i \mid \doo(X^e_{m_i} = x^{\pi}_{m_i}), \Pa^{M}(M_i), z, S=1) \right] \\  
& \quad \quad p( z \setminus z^T \mid z^T, S=1) p(z^T) \\
&=^{(4)} \sum_{z, m} p(y \mid \doo(X^e = x^{\pi}), m, z, S=1)  \left[ \prod_{i=1}^{ |M| } p(m_i \mid  x \cap \Pa_{m_i}^{\pi}, x' \cap  \Pa_{m_i}^{\bar{\pi}}, \Pa^{M}(M_i), z, S=1) \right] \\  
& \quad \quad p( z \setminus z^T \mid z^T, S=1) p(z^T) \\
&=^{(5)} \sum_{z, m} p(y \mid \doo(X^y = x^{\pi}_{y}), m, z, S=1)  \left[ \prod_{i=1}^{ |M| } p(m_i \mid  x \cap \Pa_{m_i}^{\pi}, x' \cap  \Pa_{m_i}^{\bar{\pi}}, \Pa^{M}(M_i), z, S=1) \right]  \\  
& \quad \quad p( z \setminus z^T \mid z^T, S=1) p(z^T) \\
&=^{(6)} \sum_{z, m} p(y \mid x \cap \Pa_y^{\pi}, x' \cap \Pa_y^{\bar{\pi}}, m, z, S=1)  \left[ \prod_{i=1}^{ |M| } p(m_i \mid  x \cap \Pa_{m_i}^{\pi}, x' \cap  \Pa_{m_i}^{\bar{\pi}}, \Pa^{M}(M_i), z, S=1) \right]  \\  
& \quad \quad p( z \setminus z^T \mid z^T, S=1) p(z^T)
\end{align*}

The equality $=^{(1)}$ follows from chain rule factorization.

The equality $=^{(2)}$ follows from application of Rule 1 of do-calculus. In order for this equality to hold, we require $M_i \perp \{M_1, ..., M_{i - 1}\} \setminus \Pa^{M}(M_i) \mid \Pa^{M}(M_i), Z, S$ in $G^e_{\underline{X^e}}$, the graph with edges out of $X^e$ removed. Any candidate m-connecting path between some $M_k \in \{M_1, ..., M_{i - 1}\} \setminus \Pa^{M}(M_i)$ and $M_i$ cannot be a causal path from $M_k$ to $M_i$ as this would be blocked by $\Pa^{M}(M_i)$. Suppose there were a noncausal path from $M_k$ to $M_i$ not blocked by $\{ Z, S\}$. If we concatenate this path between $M_k$ and $M_i$ with the causal path from $M_i$ to $Y$, then Lemma \ref{prop:med-exposure-outcome-indep} (c) is violated for $M_k$, therefore no such m-connecting path given $\{Z,S\}$ can exist. Suppose additionally conditioning on $\Pa^{M}(M_i)$ opens an m-connecting path between $M_k$ and $M_i$. That could occur only if some $M_j \in \Pa^{M}(M_i)$ is a collider on an unblocked path from $M_k$ to $M_i$, but then there would be an unblocked backdoor path from $M_j$ to $Y$ violating Lemma \ref{prop:med-exposure-outcome-indep} (c).

The equality $=^{(3)}$ comes from Rule 3 and removes the intervention on $X^e \setminus X^e_{m_i}$ for $p(m_i \mid \doo(X^e_{m_i} = x^{\pi}_{m_i}), \Pa^{M}(M_i), z, S=1)$ for all $i$. We require $M_i \perp X^e \setminus X^e_{m_i} \mid \Pa^{M}(M_i), Z, S$ in $G^e_{\overline{X^e_{m_i}}, \overline{X^e \setminus (X^e \cap \An(\Pa(M_i)))}}$, the graph obtained by removing edges into $X^e_{m_i}$ and removing edges into any extended node that is not an ancestor of $\Pa(M_i)$. Since $X^e_{-\An(M_i)}$ is only m-connected to $X$ and mediators that are children of $X^e_{-\An(M_i)}$, any m-connecting path between $M_i$ and $X^e \setminus X^e_{m_i}$ must go through (1) $X$ or (2) mediators that are children of $X^e_{-\An(M_i)}$. (1): any m-connecting path connecting $M_i$ and $X$ is blocked by Lemma \ref{prop:med-exposure-outcome-indep} (h). (2): paths using the edges $X \rightarrow X^e_{-\An(M_i)}$ are blocked by Lemma \ref{prop:med-exposure-outcome-indep} (f):  if $M_i$ has an unblocked path to some $M_k \in \Ch(X^e_{-\An(M_i)})$, then concatenating the path from $M_i$ and $M_k$ with the path from $M_i$ and $Y$ violates Lemma \ref{prop:med-exposure-outcome-indep} (f).

The equality $=^{(4)}$ comes from Rule 2. We require $M_i \perp X^e_{m_i} \mid \Pa^{M}(M_i), Z, S$ in $G^e_{\underline{X^e_{m_i}}}$, the graph obtained by removing edges out of $X^e_{m_i}$. $G^e_{\underline{X^e_{m_i}}}$ differs from $G^{e, pbd}_{X^e, M_i}$ as the latter removes edges out of $X^e_{\An(M_i)} = X^e \cap \An(M_i)$ instead of just $X^e_{m_i}$. Any candidate m-connecting pathway between $M_i$ and $X$ in $G^e_{\underline{X^e_{m_i}}}$ that does not intersect some $X^e_j \in X^e_{\An(M_i)}$ violates Lemma \ref{prop:med-exposure-outcome-indep} (h). Causal paths m-connecting $X^e_j$ and $M_i$ are blocked by $\Pa^{M}(M_i)$. Noncausal paths between $X^e_j$ and $M_i$ are blocked by Lemma \ref{prop:med-exposure-outcome-indep} (f):  if $M_i$ has an unblocked path to some $M_j = \Ch(X^e_j)$, then concatenating the path from $M_i$ and $M_j$ with the path from $M_i$ and $Y$ violates Lemma \ref{prop:med-exposure-outcome-indep} (f).



The equality $=^{(5)}$ follows by Rule 3. We require $Y \perp X^e \setminus X^e_y \mid X^e_y, M, Z, S$ in $G^e_{\underline{X^e_y}}$, the graph obtained by removing edges out of $X^e_y$. $X^e_y$ cannot be a collider by construction so conditioning on it cannot change its collider/non-collider status along any paths. Noncausal paths between $Y$ and $X$ that do not intersect $M$ are blocked by Lemma \ref{prop:med-exposure-outcome-indep} (b). Causal paths from $X^e \setminus X^e_y $ to $Y$ in $G^e_{\underline{X^e_y}}$ are blocked by $M$ which intercept all possible paths from $X^e \setminus X^y$ to $Y$. Any $M_i \in M$ that would act as a collider and open an m-connecting path between $X^e \setminus X^e_y$ and $Y$ would violate Lemma \ref{prop:med-exposure-outcome-indep} (d) for $M_i$.

The equality $=^{(6)}$ follows by Rule 2. We require $Y \perp X^e_y \mid M, Z, S $ hold in $G^e_{\underline{X^e_y}}$, the graph obtained by removing edges out of $X^e_y$. Removing the edge out of $X^e_y$ leaves no way for $Y$ to be m-connected to the $X^e_y$ as $\{ Z, S \}$ block backdoor paths between $Y$ and $X$ (the only other node connected to $X^e_y$) by Lemma \ref{prop:med-exposure-outcome-indep} (a). Any $M_i$ that would act as a collider and open a path between $X^e_y$ and $Y$ would violate Lemma \ref{prop:med-exposure-outcome-indep} (d) for that $M_i$.

\end{appendices}

\bibliographystyle{abbrvnat}
\bibliography{ref}

\end{document}